\documentclass{article}
\usepackage{ifxetex}
\usepackage{subfigure}
\usepackage{bm}
\usepackage{color}
\usepackage{graphicx}
\ifxetex
\usepackage{fontspec}
\usepackage{amsmath, amssymb}
\usepackage[squaren,Gray,sistyle]{SIunits}
\else
\usepackage [french]{babel}
\usepackage[T1]{fontenc}
\usepackage[latin1]{inputenc}
\usepackage{xspace}
\fi
\usepackage{textcomp}
\usepackage[hmargin=3.5cm,vmargin=2.5cm]{geometry}
\usepackage{amssymb}
\usepackage{fancyhdr}

\def\ie{i.e.,~}

\def\be{\begin{equation}}
\def\ee{  \end{equation}}
\def\ba{\begin{array}}
\def\ea{  \end{array}}

\begin{document}

\vspace{25pt}
\begin{center}
\baselineskip=13pt {\huge{}\textbf{Optimisation et identification intégrée à la
corrélation d'images de paramètres de lois élastoplastiques}}

{\huge{}\textbf{}}

\vspace{13pt}
{\large{}\textbf{M. BERTIN}}{\large{}\textbf{, F. HILD}}{\large{}\textbf{, S. ROUX}}

\vspace{13pt}
\normalsize{LMT (ENS Cachan, CNRS, Université Paris Saclay)}\\
 \normalsize{61, avenue du Président Wilson}\\
\normalsize{94235 Cachan, France}\\
\vspace{5pt}
e-mail: \{morgan.bertin; francois.hild; stephane.roux\}@lmt.ens-cachan.fr,
\end{center}

\vspace{13pt}
\baselineskip=13pt
\leftskip=0pt
{\Large{}\textbf{Résumé :}}

\vspace{13pt} {\sl L'identification de paramètres d'une loi élastique et de deux lois de comportement élastoplastique à écrouissage cinématique intégrée à la corrélation d'images (CINI) est menée sur une éprouvette biaxiale cruciforme. La géométrie est optimisée spécifiquement pour l'identification d'une loi élastique linéaire.
%la loi que l'on cherche à
%identifier. Celle-ci est conduite pour une loi élastique linéaire.
Cette optimisation s'exprime dans le formalisme de la mesure des champs de déplacement et considère l'ensemble de l'essai depuis l'acquisition d'images et la mesure d'efforts jusqu'à l'estimation des paramètres constitutifs. Par conséquent, l'ensemble des incertitudes existantes est pris en compte dans la définition du protocole expérimental.}

\vspace{13pt} {\Large{}\textbf{Abstract:}}

\vspace{13pt} {\sl Integrated digital image correlation (IDIC) is applied to identify the constitutive parameters of an elastic and two elastoplastic laws with kinematic hardening. An experiment is conducted on a cruciform specimen in a biaxial setup. The geometry of the specimen is optimized to allow for the least uncertainty in the identification of parameters of linear elastic law. The optimization takes into account the complete experimental and identification processes. Thus, full field measurements, direct numerical simulations and identification procedures accounting for uncertainties, all contribute at their best to the resulting experimental protocol.}

\vspace{13pt} {\large{}\textbf{Mots clefs~: Corrélation d'images,
Elastoplasticité, Identification}}

\vspace{13pt}
{\Large{}\textbf{1 Introduction}}
\vspace{13pt}\\
L'identification rapide et robuste des propriétés mécaniques sous chargements monotone et cyclique est un enjeu majeur pour la conception mécanique. Par exemple, l'objectif de réaliser des structures plus légères et plus sûres, conçues dans des délais toujours plus courts, nécessite une description plus précise et plus large des comportements des matériaux retenus. Un des premiers exemples concerne la caractérisation du comportement mécanique de tôles d'acier dont l'orientation de la structure vis-à-vis de la direction de laminage~\cite{Hill_1948a} affecte le comportement. Les enjeux sont d'identifier les lois et les paramètres de manière robuste et rapide. La technique d'identification intégrée à la corrélation d'images ({CINI}) apparaît être une solution capable de répondre à ces objectifs. Cette dernière a déjà été mise en {\oe}uvre pour la caractérisation de différents comportements mécaniques~\cite{Lecompte_2007a,Avril_2008b,Gras_2012,Mathieu_2014b} et fera l'objet d'une brève présentation. Néanmoins et afin de tirer le meilleur parti de celle-ci, une technique d'optimisation est proposée et dont une application est faite pour les paramètres d'un loi élastique linéaire et isotrope sur une éprouvette cruciforme. Enfin, l'identification des paramètres d'une loi linéaire élastique et de deux lois élastoplastiques à écrouissage cinématique, $i)$ linéaire et $ii)$ non-linéaire est réalisée.

\vspace{13pt} {\Large{}\textbf{2 Outils}} \vspace{13pt}

La corrélation d'images est conduite entre une image de la configuration de référence caractérisée par son niveau de gris sur chaque pixel $f(\bm{x})$ et une série d'images dans des configurations déformées $g(\bm{x})$~\cite{Sutton1983133,Besnard_2006a}. L'hypothèse de la conservation des niveaux de gris au cours du temps permet de poser comme seules inconnues du problème les composantes du champ de déplacement ($\bm{u}(\bm{x})$) entre ces images.  Le problème à résoudre s'exprime par la minimisation d'une fonctionnelle $\chi^2_f=1/N_t\sum_t \chi^2_f(t)$ qui est la différence quadratique moyenne sur l'ensemble de la région étudiée entre les images $f$ et $g$ corrigées du champ de déplacement mesuré $\bm{u}(\bm{x},t)$
\begin{equation}
    \chi^2_{f}(t) = \frac{1}{2 \gamma^2_f N_{\Omega}} \sum_{\Omega} ((g(\bm{x}+\bm{u}(\bm{x},t),t)-f(\bm{x}))^2
    \label{eq:eq1}
\end{equation}
où $N_{\Omega}$ est le nombre de pixels dans la région sur laquelle la corrélation d'images est menée, et $N_t$ le nombre d'images acquises au cours de l'essai. Pour l'identification intégrée, les inconnues sont les paramètres inconnus de la loi de comportement choisie. Le problème revient à trouver le meilleur jeu de paramètres qui minimise globalement la fonctionnelle $\chi^2_{f}$ par rapport aux champs de déplacement appartenant au sous-espace engendré par les champs de sensibilité aux paramètres~\cite{Mathieu_2014b} % le vecteur de résidus $\{\bm{b}\}$
    \begin{equation}
    \{\delta \bm p\}^{}=\frac{1}{2\gamma^2_f}
    [{\bm M}]^{-1}_{IDIC} [\bm{S_U}]^t  \{\bm{b}\}
    \end{equation}
où $[\bm{M}]_{IDIC}=1/(2\gamma^2_f)[\bm S_U]^{t} [\bm{M}][\bm S_U]$ est le hessien cinématique, $[\bm M]$ la matrice de corrélation d'images pour une discrétisation choisie, $[\bm S_U]$ est la matrice contenant toutes les composantes des champs de sensibilité~\cite{Lecompte_2007a} vis-à-vis des paramètres recherchés ($[\bm S_U]=[\partial \bm{u} / \partial \bm{p}]$), $\{\bm{p}\}$ est le vecteur contenant les paramètres recherchés et enfin $\{\bm{b}\}=(f-g)(\bm{\nabla} f)$. Par ailleurs, les efforts mesurés peuvent aussi servir à l'identification. L'écart entre les valeurs mesurées et simulées s'exprime dans la fonctionnelle suivante que l'on cherche à minimiser
\begin{equation}
    \label{eq:chi_F}
    \chi^2_{F}=\frac{1}{N_{F}N_t \gamma^2_F}\{\bm F_m-\bm F_c\}^t \{\bm F_m-\bm F_c\}
\end{equation}
où $N_F$ est le nombre de données mesurées sur les efforts à chaque pas de temps, et $N_t$ le nombre de temps de mesure. Pour cette seule fonctionnelle, la linéarisation du problème revient à résoudre successivement jusqu'à convergence
\begin{equation}
    \{\delta \bm p\}^{}=\frac{1}{\gamma^2_F}[\bm H]_F^{-1} [\bm{S_F}]^{t}\{\bm F_m-\bm F^{}_c\}
\end{equation}
où $[\bm{H}]_F=1/\gamma^2_F [\bm S_{\bm F}]^{t}[\bm S_{\bm F}]$ est le hessien statique, $[\bm{S}_F]$ est la matrice contenant les sensibilités en effort, $\{\bm{F_m}\}$ et $\{\bm{F_c}\}$ sont respectivement les efforts mesurés et calculés aux différents instants considérés.

Les deux types de mesure sont combinés en normant chacune des fonctionnelles par l'amplitude de leurs incertitudes respectives. Le poids spécifique provient d'une formulation bayésienne. L'hypothèse sous-jacente est l'existence d'un bruit blanc gaussien, i.e., non corrélé dans le temps. Ainsi, si seul le bruit est présent (aucune erreur de modèle) la valeur attendue de $\chi^2_{}$ est 1. Cette dernière s'exprime par
\begin{equation}
    \chi^2_{\bm{}}=
    \frac{N_{\Omega}}{N_{\Omega}+N_{F}} \chi^2_{f}
    +\frac{N_{F}}{N_{\Omega}+N_{F}} \chi^2_{F}
    \label{eq:chi_I}
\end{equation}
%
%où $N_f$ est le nombre d'inconnues cinématiques, i.e., 2 fois le
%nombre de pixel. $\{\delta \bm p\}^{(i)}$ est le vecteur contenant
%les incréments de variation des paramètres inconnus à l'itération
%$i$.
La matrice de covariance des paramètres identifiés s'écrit
alors ($[\bm{H}]_{}=[\bm{M}]_{IDIC}+[\bm{H}]_{F}$)
\begin{equation}
    [\bm{C^{}_p}]
    =\langle \{\delta \bm p\} \otimes \{\delta \bm p\} \rangle
    =[\bm{H}]_{}^{-1}
\end{equation}

\vspace{13pt}{\Large{}\textbf{3 Loi de comportement}}\vspace{13pt}

L'identification est réalisée pour une loi élastique linéaire
($A$) et deux lois élastoplastiques, la première ($B$) à
écrouissage cinématique linéaire et la seconde ($C$) à écrouissage
cinématique de forme exponentielle. Plusieurs auteurs ont déjà
proposé des études sur l'identification des paramètres de loi de
comportement élastoplastique~\cite{lecompte_2007_th,Rthor201373}.
La déformation totale est séparée en une contribution élastique et
plastique $\bm\epsilon_{tot}=\bm\epsilon_{el} + \bm\epsilon_{pl}$
où la loi élastique est isotrope et linéaire telle que $\bm\epsilon_{el}=\frac{1+\nu}{E} \bm{\sigma} - \frac{\nu}{E} \mbox{tr}(\bm{\sigma}) \bm{I}$. $\nu$ et $E$ sont respectivement le coefficient de Poisson et le module d'Young et $\bm{I}$ est le tenseur identité. Le seuil de plasticité est défini par la contrainte équivalente de Von Mises
\begin{equation}
f(\bm{\sigma}-\bm{\chi})=\sqrt{\frac{3}{2} ( \bm{\sigma}-\bm{X})^{d}:(\bm{\sigma}-\bm{X})^{d})} - \sigma_0=0
\end{equation}
où $\sigma_0$ est la limite d'élasticité macroscopique du matériau et $\bullet^d=\bullet - 1/3 \mbox{tr}(\bullet) \bm I$ est la partie déviatorique du tenseur $\bullet$. Dans le cadre d'une modélisation linéaire de l'écrouissage, la variable d'écrouissage cinématique $\dot{\bm{X}}$ s'exprime par $ \dot{\bm{X}} = \frac{2}{3}C \dot{\bm \epsilon}_{pl} $. Pour la seconde loi à écrouissage cinématique non-linéaire de forme exponentielle, son expression est $ \dot{\bm{X}} = \frac{2}{3}C \dot{\bm \epsilon}_{pl} -c\bm{X}\dot{p}$ où $\dot p$ est le taux de déformation plastique cumulée \cite{Frederick_1966,Chaboche_1989}. Enfin, les paramètres sont exprimés suivant une échelle logarithmique qui permet de privilégier les incertitudes relatives. La paramétrisation retenue s'exprime selon $\{\bm{q}\}=(\log(\{\bm p/\bm p_0\})$ où $\{\bm p\}$ est le jeu des paramètres recherchés  $\{\bm p\} = (E,\nu,\sigma_0,C,\gamma)$.

\vspace{13pt}{\Large{}\textbf{4 Optimisation}}\vspace{13pt}

Il n'est pas aisé de formuler un critère d'optimisation universel lorsque plusieurs paramètres sont en jeu. Dans le cas ici présenté, l'optimisation est naturellement multiparamétrique et un critère doit être formulé. La stratégie choisie est de viser l'incertitude la plus faible pour le paramètre le plus sensible au bruit. L'optimisation consiste donc à minimiser la plus grande des valeurs propres associées à $[\bm C_p]$ afin de minimiser l'influence du bruit sur les paramètres associés. Ceci est équivalent à maximiser la plus faible valeur propre de $[\bm H]$. Les paramètres matériaux correspondent à un acier inoxydable à durcissement structural 17-7 PH~\cite{ak_steel_17_7PH} ($E=200$ GPa, $\nu=0.3$ et $\sigma_0=1300$ MPa). A cette étape, une hypothèse de champ moyen est exploitée du fait de la méconnaissance {\it a priori} du mouchetis.  Elle permet d'exprimer la matrice de correlation comme
\begin{equation}
    M_{ij}\approx \frac{1}{2} G_f^2  \sum_{\Omega} \bm \psi_i(\bm{x}) \cdot \bm \psi_j(\bm{x})
    \label{eq:M_DIC_mean_field}
\end{equation}
où $G_f^2=\langle (\bm \nabla f)^2\rangle$ est le gradient de niveau de gris moyen et ($\psi_{i}$,$\psi_{j}$) sont les fonctions de forme choisies. L'incertitude des efforts provient de deux sources indépendantes, la première est l'incertitude des capteurs de force (dépendant du niveau d'effort). La seconde concerne les conditions aux limites de type Dirichlet (obtenues à partir d'une solution initiale du champ de déplacement à l'aide de la CIN). En effet, des niveaux de déplacement affectés du bruit d'acquisition vont conduire à des efforts simulés erronés. Le modèle d'incertitude global retenu s'exprime selon
%
%\begin{figure}[h!]
%\begin{center}
%\includegraphics[scale=0.5]{loads.eps}
%\end{center}
%\caption{Mesure de la variance de l'effort évaluée pour un
%chargement statique en fonction du niveau d'effort imposé. Une loi
%quadratique est identifiée sur les valeurs expérimentales qui
%tiennent compte des incertitudes des cellules d'efforts et des
%conditions aux limites issues de la CIN}
%\label{fig:load_uncertainty}
%\end{figure}
%
\begin{equation}
    \gamma^2_{\bm F}=\rho_1^2 |\bm F|^2+\rho_0^2
\end{equation}
où $\rho_1$ est l'écart type de l'incertitude qui dépend de l'effort imposé et $\rho_0$ une valeur constante correspondant à l'erreur due aux conditions aux limites imposées issues de la {CIN}. Cette dernière est évaluée à chaque itération et tient compte de l'évolution des valeurs des paramètres. Le tableau~\ref{tab:resolution} réunit les valeurs d'incertitudes retenues. L'historique de chargement est non proportionnel et impose dans un premier temps une amplitude de déplacement $d_1=d^*$ suivant la direction $\bm e_1$ en maintenant le déplacement suivant $\bm e_2$ à $d_2=0$. Une fois celle-ci atteinte, le déplacement est imposé suivant $\bm e_2$ jusqu'à la même amplitude $d_2=d^*$. Puis la décharge est équibiaxiale jusqu'à $d_1=d_2=0$. Dans le plan $(d_1,d_2)$, le trajet suit la forme d'un triangle, d'où sa dénomination \og chargement triangle\fg.
\begin{table}[h!]
    \centering
    \caption{Résolution et incertitude des déplacements
    et efforts mesurés. Les valeurs sont évaluées à partir
    de cas expérimentaux. La taille du pixel correspond
    à $a=13.5$~{\textmu}m}%\SR{L'unité de $\rho_0$ est le Newton.}}
    \vspace{5pt}
    \begin{tabular}{|l|c|c|c|c|c|}
    \hline
    Quantité & $\rho_0$ & $\rho_1$ &  $\gamma_f$
    & $G_f$ & $a$ \\
    \hline
    Valeur & 2.5~N & 2.9$\times10^{-4}$ &  280~niveaux de gris
    & 3800~niveaux de gris/pixel & $13.5$~{\textmu}m\\
    \hline
    \end{tabular}
    \label{tab:resolution}
    \end{table}

\begin{figure}[h!]
        \centering
 \subfigure{\includegraphics[width=0.4\textwidth]{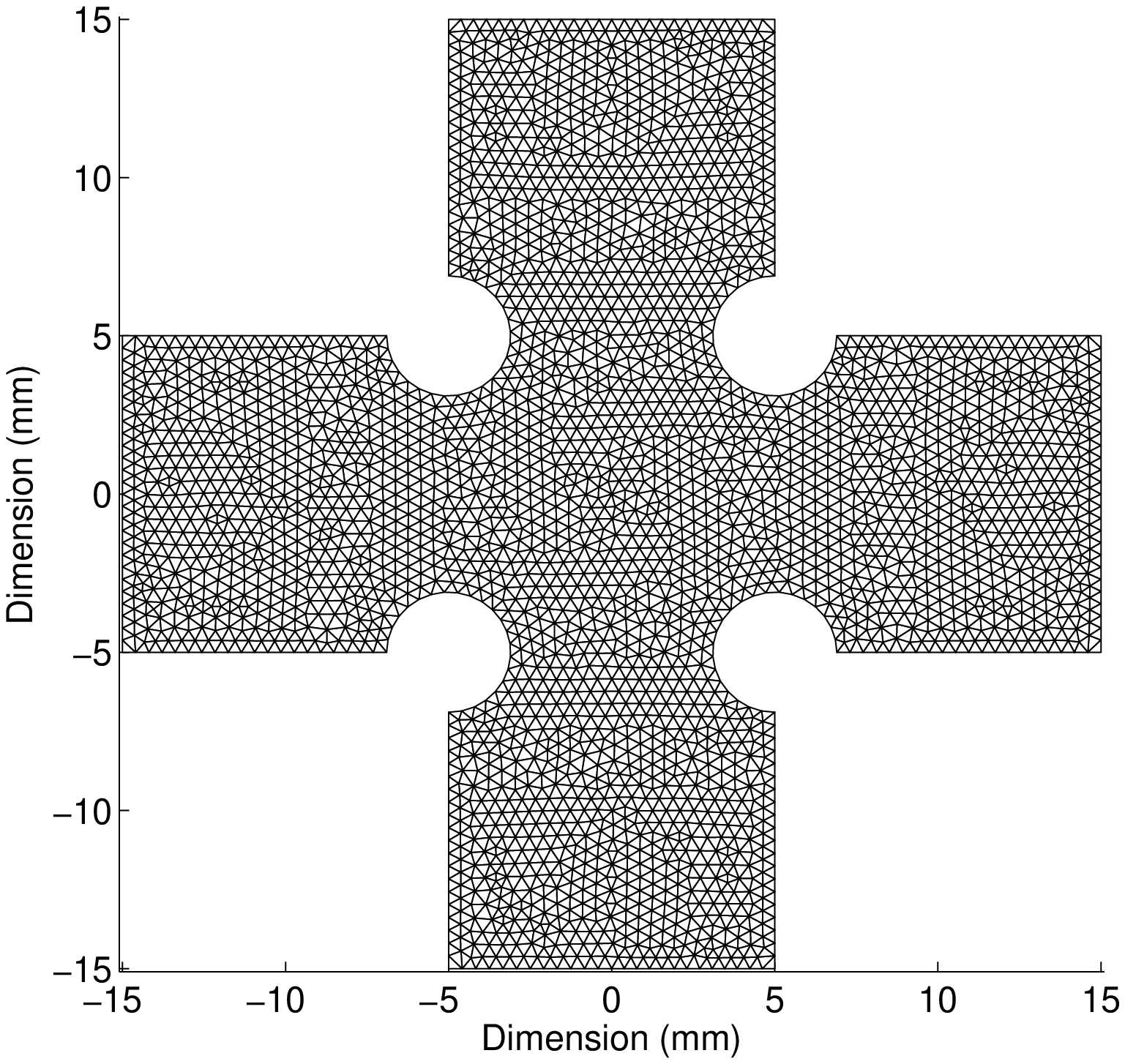}}
 \subfigure{\includegraphics[width=0.46\textwidth]{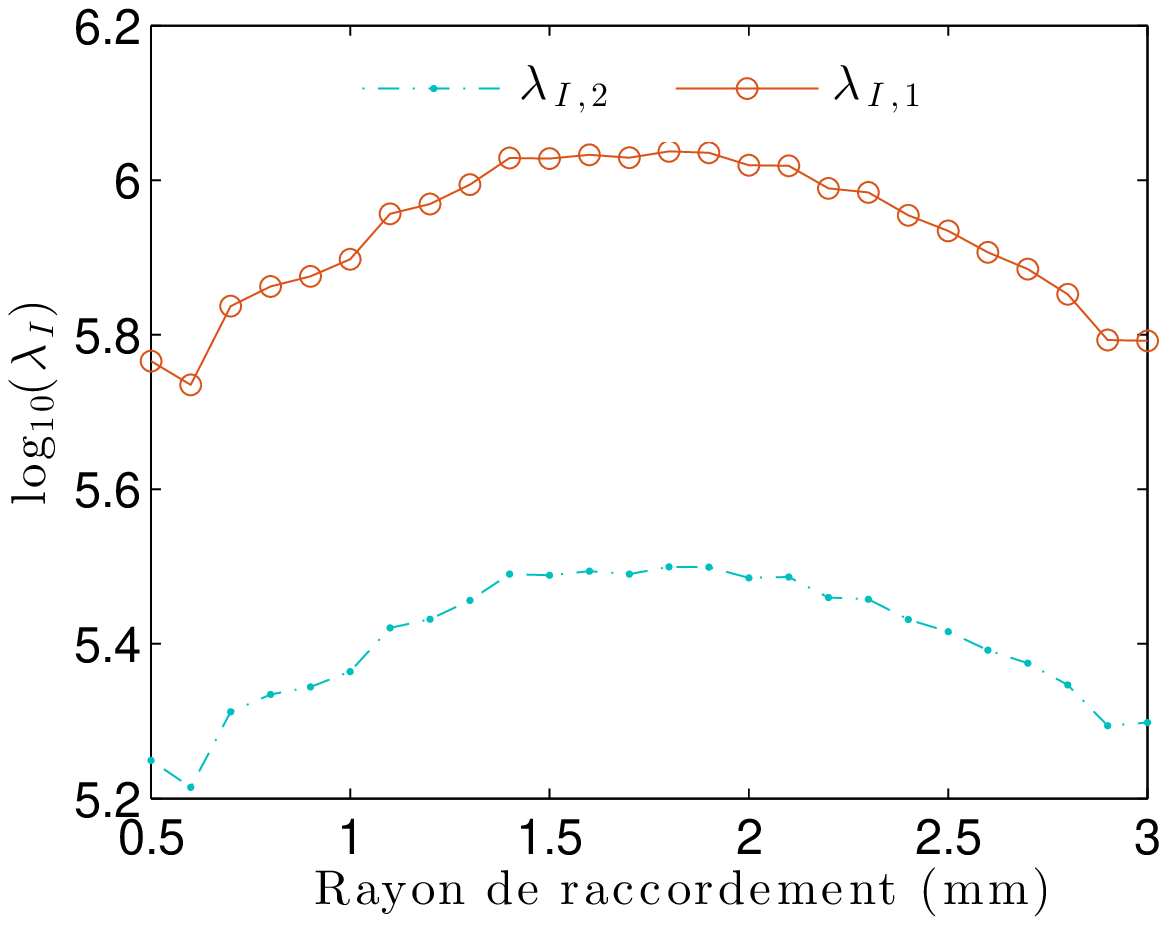}}
       \caption{(a) Maillage et géométrie de l'éprouvette biaxiale à optimiser
       pour un rayon de raccordement égal à $r=2$~mm.
       (b) Valeur propre $\lambda_{I}$ de $[\bm H]_{}$
       pour le jeu de paramètre $(\log(E/E_0),\log(\nu/\nu_0))$}
       \label{fig:opt}
\end{figure}

Le paramètre géométrique optimisé est le rayon du congé de raccordement reliant les bras de l'éprouvette. La figure~\ref{fig:opt}(a) présente une des géométries ainsi que le maillage associé. Le rayon de raccordement présenté Figure~\ref{fig:opt}(a) est $r=2$~mm. Figure~\ref{fig:opt}(b) montre les valeurs propres de la matrice hessienne $[\bm{H}]_{}$ en fonction du rayon de raccordement évalué pour l'ensemble de l'historique de chargement. L'amplitude du déplacement imposé est adaptée à chaque valeur de $r$ de telle sorte que la valeur maximale de la contrainte équivalente de Von Mises soit égale à la limite d'élasticité macroscopique du matériau. La valeur optimale du rayon est évaluée à $r=1.7$~mm, correspondant à la valeur maximale de $\lambda_{I,2}$ (figure~\ref{fig:opt}(b)).

\vspace{13pt} {\Large{}\textbf{5 Identification}}\vspace{13pt}

\begin{figure}[h!]
\begin{center}
\subfigure[]{\raisebox{16pt}{\includegraphics[width=0.35\textwidth]{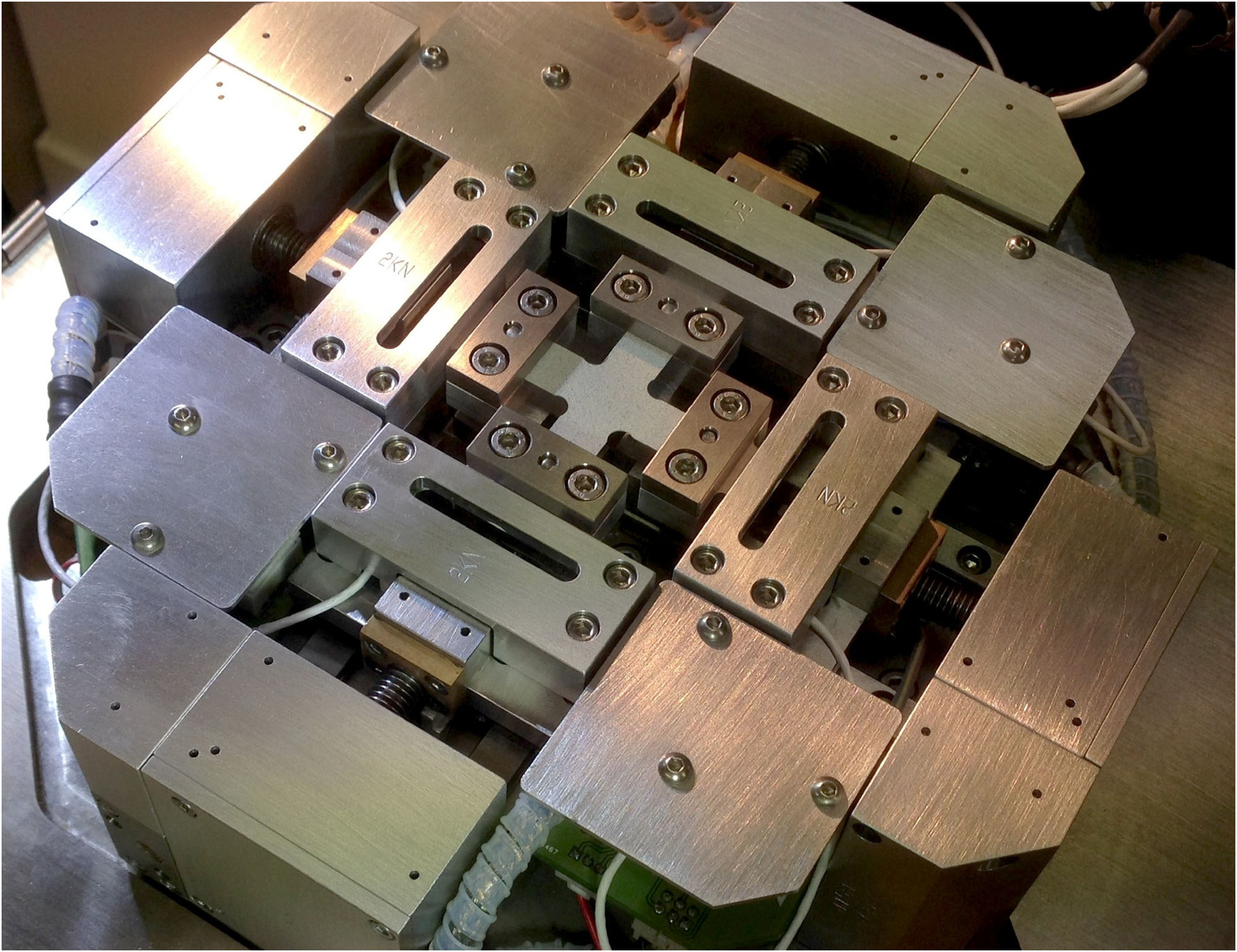}}}
\subfigure[]{\includegraphics[width=0.64\textwidth]{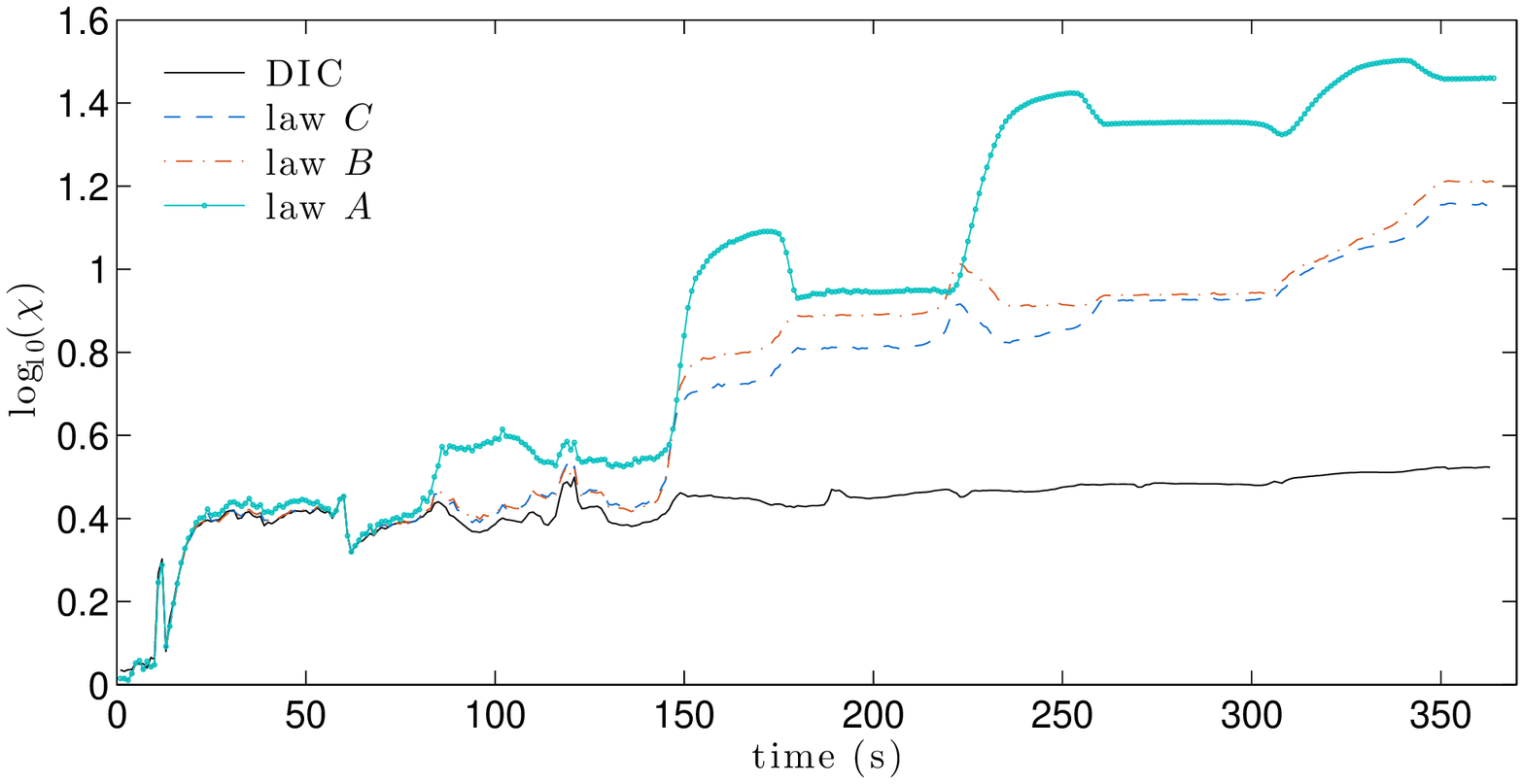}}
  \caption{(a) machine de traction compression biaxiale et (b) résidus de corrélation {CIN} (noir) et {CINI} (color).}
     \label{fig:machine}
\end{center}
\end{figure}

L'identification est réalisée sur un essai biaxial à efforts imposés et à plusieurs cycles à l'aide du dispositif expérimental présenté Figure~\ref{fig:machine}. Le matériau est l'acier inoxydable 17-7 PH avant précipitation par traitement thermique (le comportement est plus ductile). Ces caractéristiques sont différentes de celles retenues pour l'optimisation avec une déformation totale potentielle plus importante ($\epsilon \approx 35 \%$) mais avec une limite élastique plus faible ($\sigma_0=300$~MPa). Le chargement est aussi non proportionnel et triangulaire mais est à efforts imposés. Les images sont acquises à l'aide d'une caméra 16-bits PCO Edge et un objectif télécentrique de grossissement $\times 0.5$. Les paramètres initiaux sont ceux identifiés dans un premier temps uniquement à partir des efforts de réaction (Equation \ref{eq:chi_F}).
La Figure \ref{fig:machine}(b) montre les résidus de corrélation à convergence issus du calcul {CIN} initial (noir) et des calculs {CINI} (couleurs). Après un début d'essai pour lequel les résidus {CIN} et {CINI} sont sensiblement identiques, ceux-ci s'écartent une fois que les déformations deviennent plus importantes. Le résidu pour la loi $A$ s'écarte de la solution $CIN$ à partir du pas de temps $t=25$, tandis que les résidus de $B$ et $C$ s'écartent à partir du pas de temps $t=80$. Il apparaît à la fin de l'expérience que le résidu d'identification le plus faible est celui obtenu pour la loi $C$. Enfin, une fonctionnelle de régularisation ($\chi_R$) est associée à la fonctionnelle à minimizer ($\chi$) afin de prévenir une évolution irraisonné des paramètres dont la sensibilité est inférieure au bruit. $\chi_R$ est une fonction convexe et son minimum est égal  zéro. Celle-ci s'exprime
\begin{equation}
\chi^2_R=\{\bm q\}[\bm C^R_q]^{-1} \{\bm q\}
\end{equation}
où $[\bm C^{R}_q]$ est la matrice de covariance associée à la résolution limite choisie. Enfin et puisque $\chi_R$ et $\chi$ sont sans dimensions et dimensionnés à 1 la fonction à minimiser régulariser devient
\begin{equation}
\left([\bm H ]_{}+\lambda^*[\bm I ]\right)^{}\{\delta \bm q \}=\{\bm b \}_{}+\lambda^*\{\bm q\}
\label{eq:modified_N}
\end{equation}
où $\lambda^*$ est le paramètre de régularisation. Celui-ci est choisi de tel sorte que les paramètres dont l'influence sur les grandeurs mesurables soient inférieurs à l'incertitude de mesure ne subissent pas d'évolution. La table 2 montre les paramètres identifiés sur le premier cycle où un très faible niveau de déformation plastique est atteinte et où celui-ci n'est pas suffisant pour révéler l'influence des paramètres matériaux. Par conséquence et grâce à la régularisation leurs valeurs ne subissent pas de variation.
\begin{table}[h!]
\centering \caption{Identified parameters and identification
residuals for the three laws for the first cycle ($\gamma_p$ is
expressed in $\%$ of $p$).} \label{tab:cycle_1_res} \footnotesize
\begin{tabular}{|c|ccccccccccccc|}
\hline
loi & $\chi_{}$ & $\chi_{f}$ & $\chi_{F}$ & $E$ (GPa)& $\gamma_E$ & $\nu$ & $\gamma_{\nu}$ & $\sigma_0$(MPa) & $ \gamma_{\sigma_0}$ & $C$(GPa) & $\gamma_{C}$ & $c$ & $\gamma_{c}$ \\
    &               &            &            & 200 &            &   0.3    &                &  300       &                      & 10 &              &   10  & \\
\hline
$A_{}$ & 2.10 & 2.10 &  15 & 174 & 0.4   & 0.31 & 0.09 & --- &  --- & ---& --- & --- & ---\\
$B_{}$ & 2.09 & 2.09 & 8.9 & 195 & 0.21 & 0.297 & 0.1 & 306 & 0.25  & 10 & %3
--- & --- & ---\\
$C_{}$ & 2.09 & 2.09 & 8.9 & 195 & 0.25 & 0.297 & 0.1 & 306 & 0.3  & 10 & %11  & 10  & $5.9\times 10^5$\\
--- & 10 & --- \\
\hline
\end{tabular}
\normalsize
\end{table}

\begin{figure}[h!]
\begin{center}
\subfigure[]{\includegraphics[width=0.25\textwidth]{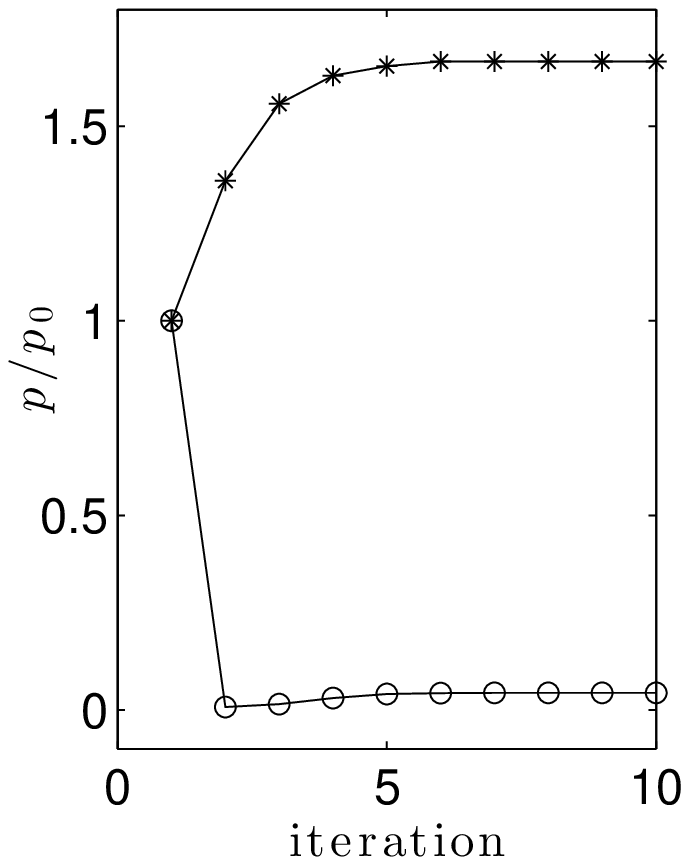}}
\subfigure[]{\includegraphics[width=0.25\textwidth]{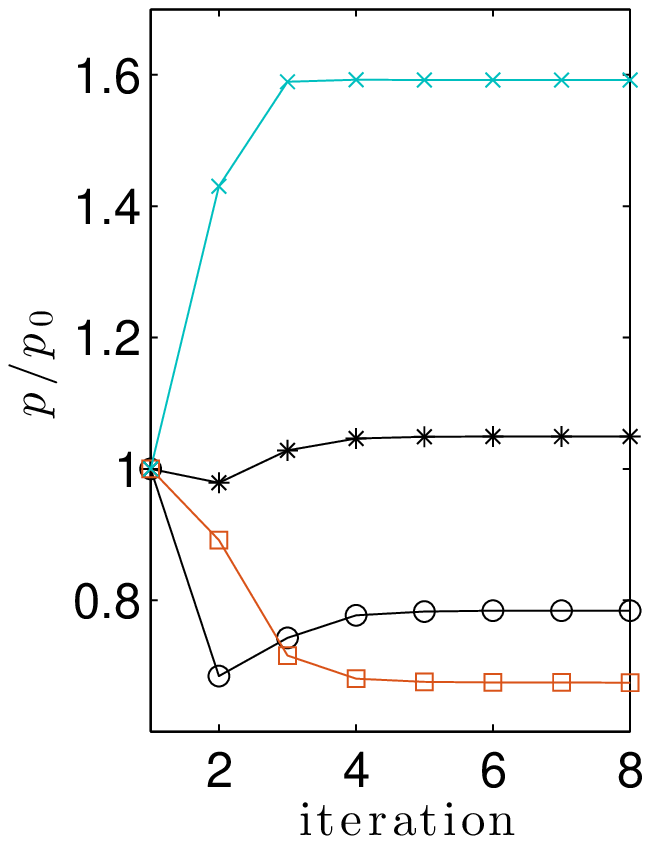}}
\subfigure[]{\includegraphics[width=0.33\textwidth]{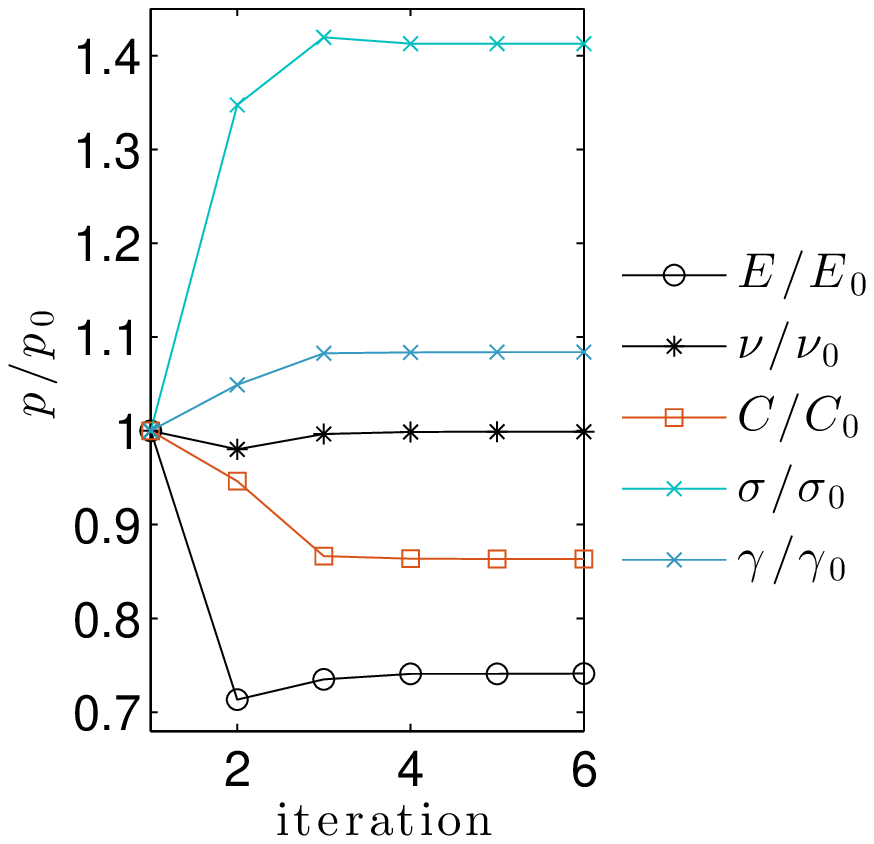}}
  \caption{Evolution des paramètres matériaux respectivement pour les lois (a) $A$ , (b) $B$ et (c) $C$.}
      \label{fig:iteration}
\end{center}
\end{figure}

 La Figure~\ref{fig:iteration} montre les évolutions de ceux-ci au fur et à mesure des itérations. Pour la loi $A$, le paramètre $\nu$ converge vers $\nu=0.50$, en effet le comportement étant plastique, il apparaît quasi-incompressible. Les autres paramètres convergent relativement rapidement pour les lois $B$ et $C$.
Le tableau~3 montre les résultats de l'identification sur l'essai complet. Le module d'Young identifié pour la loi $A$ n'est pas physique. En effet, la déformation étant principalement plastique, et la loi élastique linéaire, le module d'Young identifié est ajusté à une valeur trop basse pour rester compatible avec les efforts mesurés. Une loi élastoplastique est plus satisfaisante, et conduit à un résidu d'identification plus faible.  Pour la loi à écrouissage cinématique non-linéaire, ce résidu est de l'ordre de 6.5 fois le niveau de bruit, montrant encore une erreur de modèle, mais cependant une amélioration sensible par rapport au cas élastique. Le résidu est légèrement plus élevé pour la loi à écrouissage cinématique linéaire. %\SR{  Je ne comprends pas la phrase: Concernant l'influence du bruit d'acquisition sur les paramètres identifiés, il apparait que les paramètres modélisant l'écrouissage cinématique sont eux aussi digne de confiance.}
\begin{table}[h!]
\centering \caption{Identified parameters and identification
residuals for the three laws for the whole loading history}
\label{tab:cycle_5_res}  \footnotesize
\begin{tabular}{|c|ccccccccccccc|}
\hline
loi & $\chi_{}$ & $\chi_{f}$ & $\chi_{F}$ & $E$ (GPa)& $\gamma_E$ & $\nu$ & $\gamma_{\nu}$ & $\sigma_0$(MPa) & $ \gamma_{\sigma_0}$ & $C$(GPa) & $\gamma_{C}$ & $c$ & $\gamma_{c}$ \\
    &               &            &            & 200 &            &   0.3    &                &  300       &                      & 10 &              &   10  & \\
\hline
$A_{}$ &  15 & 14.9 & 1820 & 8.8 & 0.25 & 0.499 & 0.002 & --- & ---  & --- & ---  & ---  & --- \\
$B_{}$ & 7.1 &  7.1 &  113 & 157 & 0.15 & 0.31  & 0.04  & 480 & 0.15 & 6.7 & 0.15  & ---  & --- \\
$C_{}$ & 6.3 &  6.3 &  100 & 148 & 0.16 & 0.30  & 0.04  & 423 & 0.16 & 8.6 & 0.16 & 10.8 & 0.03\\
\hline
\end{tabular}\normalsize
\end{table}

\begin{figure}[h!]
\begin{center}
\subfigure[Loi $A$]{\includegraphics[width=0.24\textwidth]{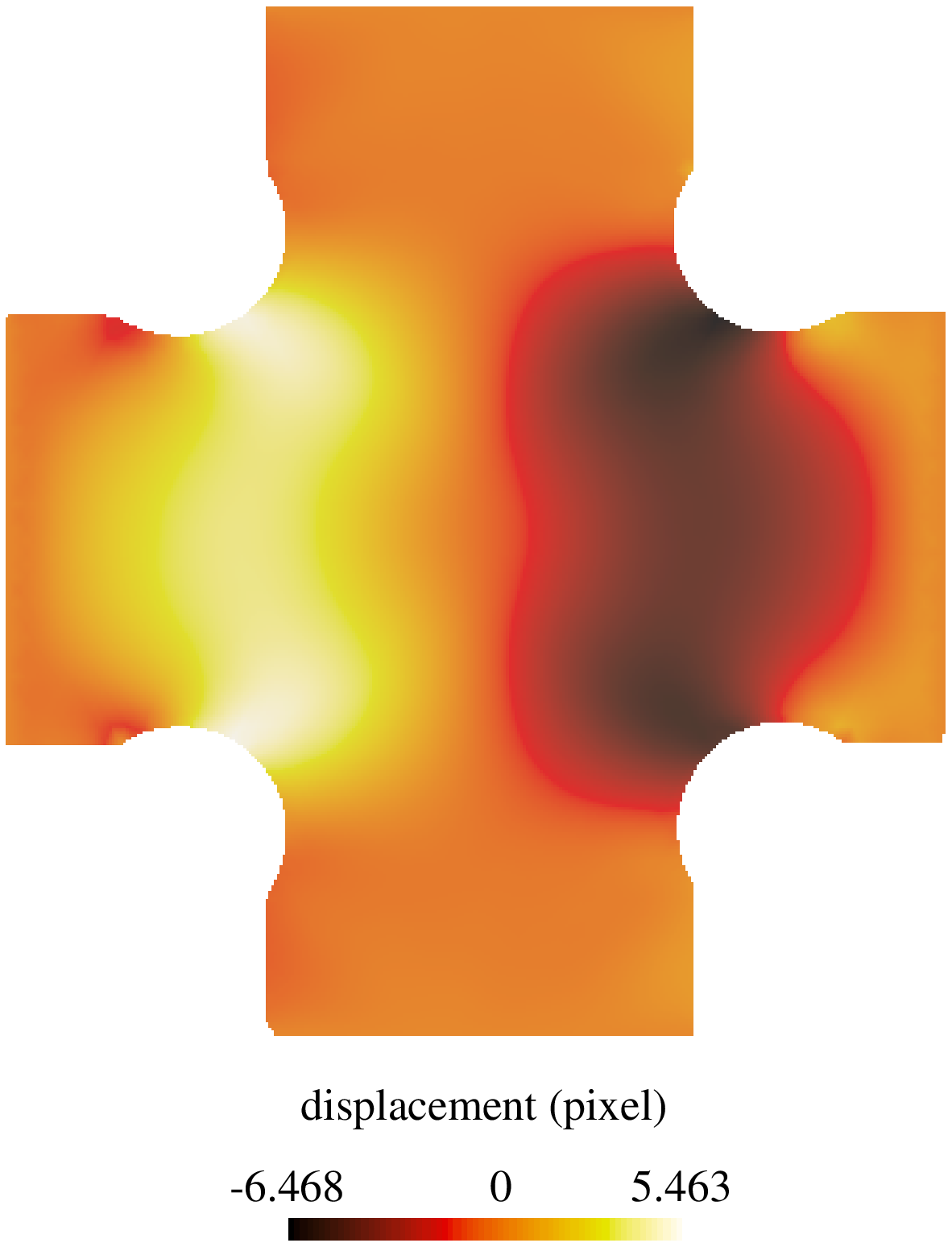}}
\subfigure[Loi $B$]{\includegraphics[width=0.24\textwidth]{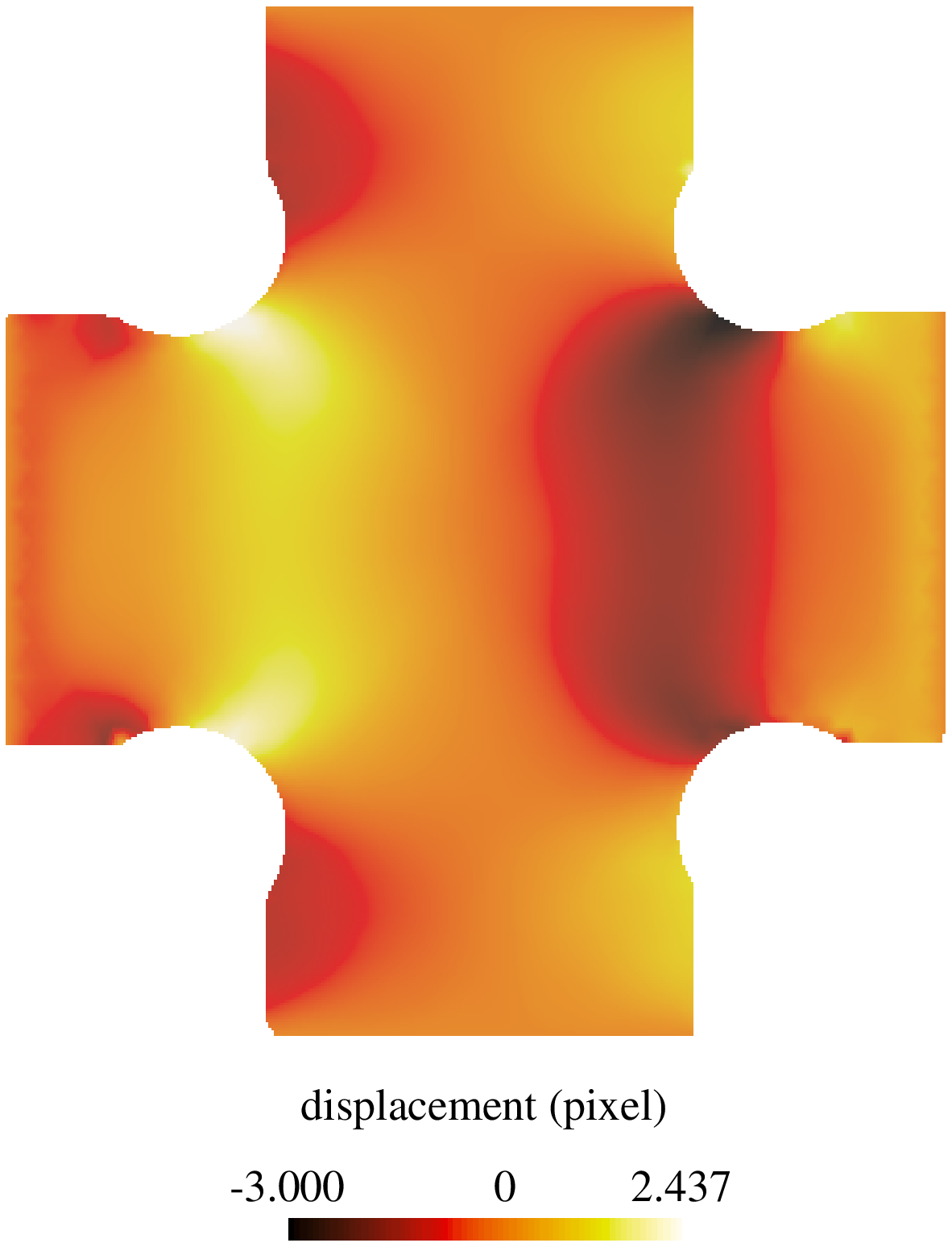}}
\subfigure[Loi $C$]{\includegraphics[width=0.24\textwidth]{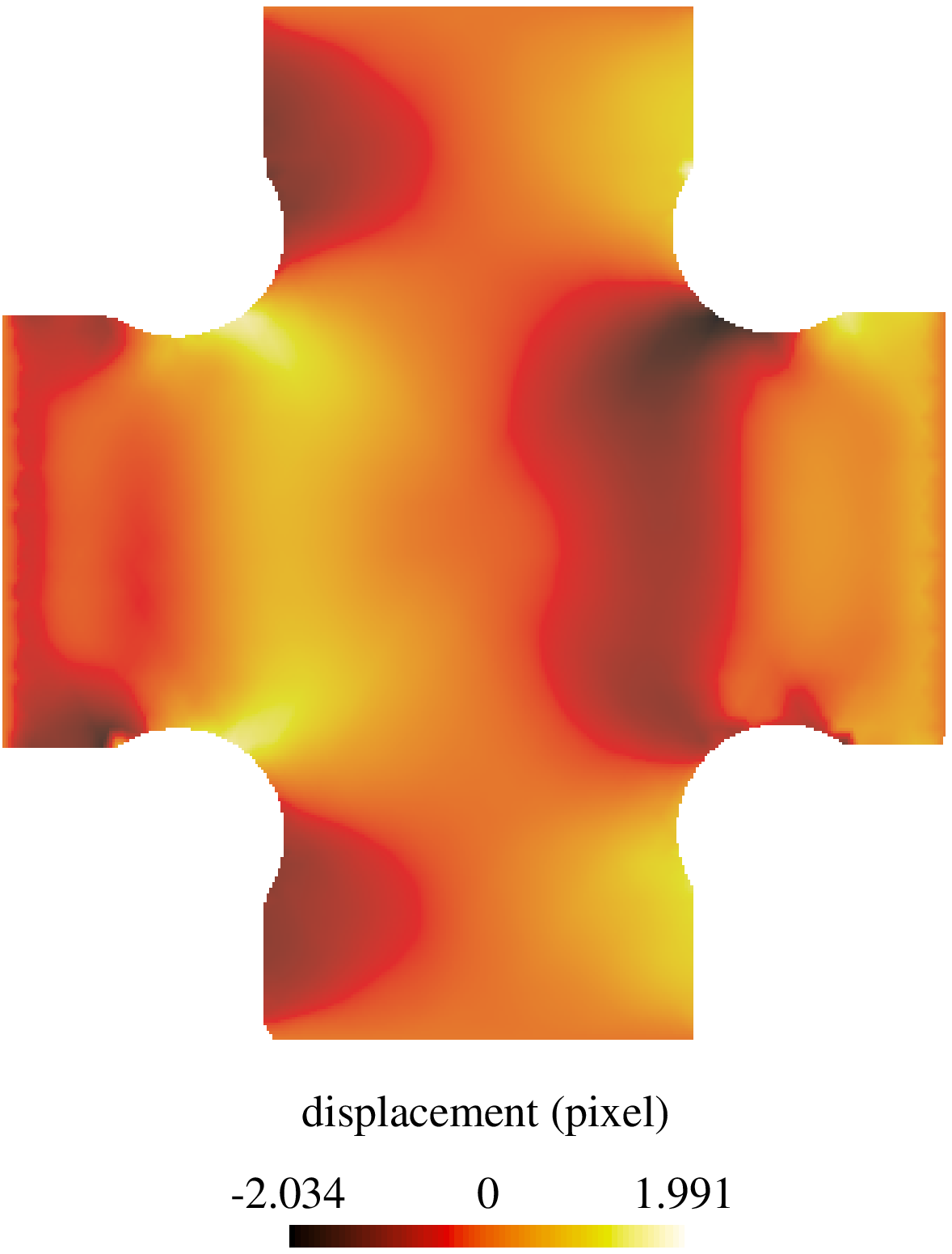}}
\subfigure[CIN]{\includegraphics[width=0.24\textwidth]{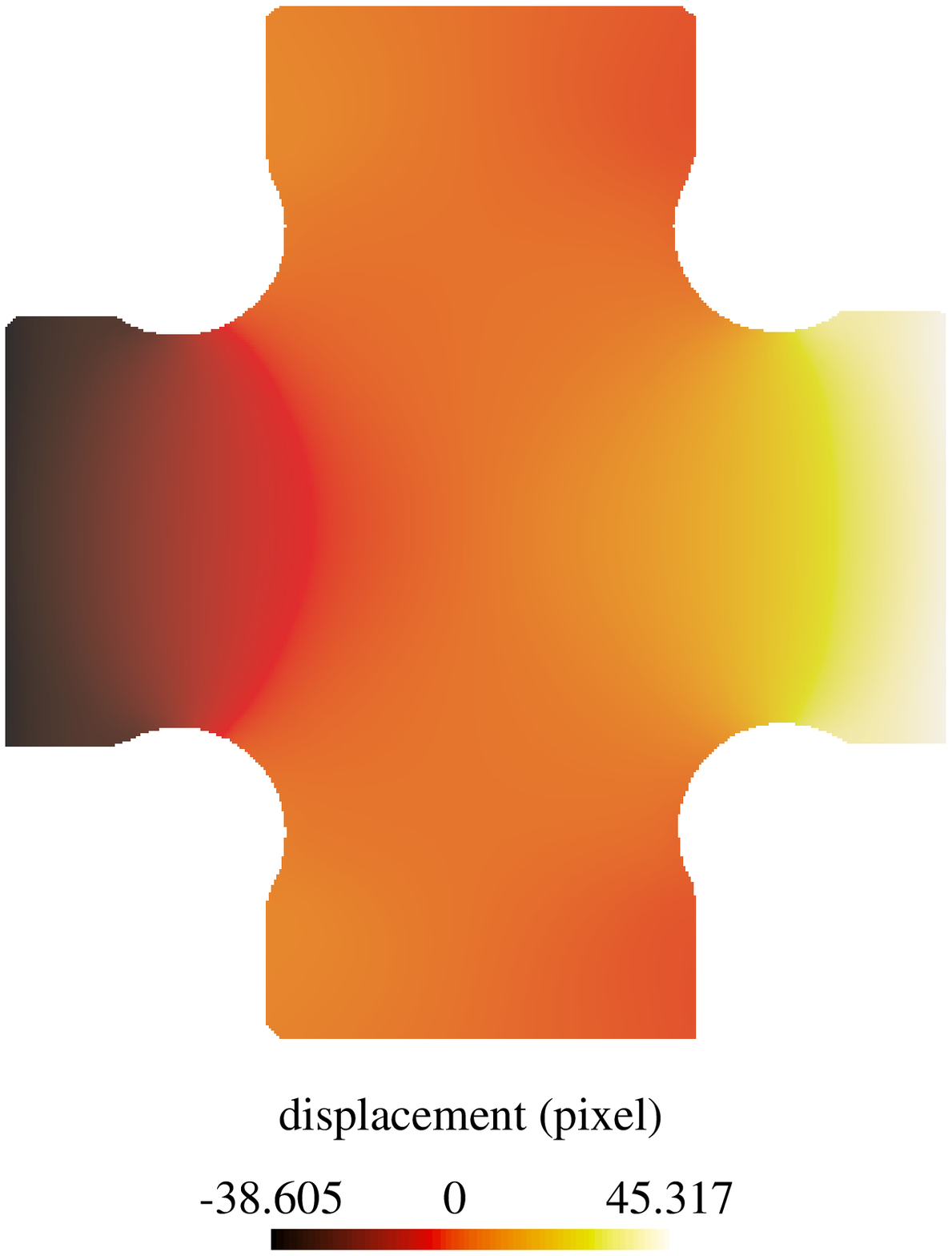}}
  \caption{(a), (b) and (c) différence entre les champs de déplacements issues de {CIN} et de {CINI} pour les lois $A$, $B$ et $C$ à la fin de l'expérience. (d) composante $U_1$ du champ de déplacement $CIN$ au même instant}
      \label{fig:residu_U}
\end{center}
\end{figure}

La figure~\ref{fig:residu_U} présente la composante  $U_1$ du résidu champ de déplacement à la fin de l'expérience, i.e., $t=350$~s, où encore $\epsilon_{max}=18.9 \%$. De la gauche vers la droite, le résidu pour la loi $A$, i.e., élastique linéaire, la loi $B$ et enfin $C$. Pour la première, le résidu correspond directement aux déformations permanentes qui ne sont pas prédits par une loi linéaire élastique. Concernant les deux autres composantes du déplacement résiduel, la loi $B$ est à un niveau faible (dynamique de déplacement réduite d'un facteur 2) et la loi $C$ améliore encore ce niveau de résidus. Malgré tout, des déformations (hors de l'incertitude de mesure) restent inexpliquées. Il y a donc matière à généraliser la forme algébrique de la loi de comportement pour mieux s'approcher des observations expérimentales.

\vspace{13pt}{\Large{}\textbf{6 Conclusion}}\vspace{13pt}\\
Une nouvelle approche d'identification optimisée a été proposée. Sa mise en {\oe}uvre sur une éprouvette biaxiale a permis d'identifier une géométrie minimisant l'influence des incertitudes de mesure sur les paramètres à identifier dans le cas d'une loi élastique linéaire. Pour cela, l'ensemble de la chaîne de mesures exploitée pour l'identification intégrée est considéré. L'optimisation est spécifiquement réalisée pour la loi dont on cherche à identifier les paramètres et dont on suppose {\it a priori} le comportement expérimental.
L'étude expérimentale est cependant effectuée dans le cas d'un trajet de chargement qui explore largement le domaine plastique. L'identification a été réalisée pour trois lois de comportement et les paramètres ont été identifiés avec un niveau d'incertitude faible. Néanmoins, le résidu d'identification, relativement élevé en comparaison du résidu {CIN}, montre clairement l'existence d'une erreur de modèle résiduelle, très forte pour une loi élastique linéaire, et bien moindre pour une loi élasto-plastique. Ceci permet de classer les modèles en fonction de leurs qualités respectives vis-à-vis des données expérimentales.
%
% Advantages of the study
%
Dans la présente étude quatre indicateurs d'erreur ont été considérés,\\
\begin{itemize}
\item Quand les analyses CIN et CINI sont employées, l'écart par rapport à l'hypothèse de conservation de niveau de gris est évalué. Il estime la qualité de mesure lorsque les images déformées sont corrigées par le champ de déplacement mesuré et comparées à l'image de référence.
\item CIN et CINI peuvent aussi être comparées en calculant les déplacements mesurés par les deux approches. Lorsque le même maillage est employé, la comparaison est directe.
\item Le résidu d'effort et un autre moyen de caractériser la qualité de l'identification. Ici, FEMU-F et CINI peuvent être comparées car les deux approches inclues les efforts dans leurs formulations respectives.
\item Finalement, la méthode d'identification intégrée CINI possède un indicateur de qualité qui considère toutes les sources d'informations utilisées dans un unique formalisme grâce à la normalisation par les variances des l'incertitudes correspondantes.
\end{itemize}
~\vspace{5pt}

L'ensemble de ces indicateurs permet à l'utilisateur d'aboutir à une identification globale mais aussi avec évaluer la qualité de l'identification et l'erreur de modèle (\ie choix d'une loi de comportement, discrétisation en éléments finis). Cette étude cherche à présenter et à construire une route pour la compréhension des phénomènes mécaniques, la conception optimisée des essais expérimentaux et l'identification pour valider mais aussi invalider les modèles étudiés. Dans des études futures, il est envisagé d'étendre cette technique à des matériaux différents tels que des polymères (comportement hyper-élastique), céramiques (comportement fragile) ou encore composites (comportement anisotrope). Enfin, des problèmes à l'échelle microscopique avec la prise en compte de la microstructure pourront être envisagés.
%
%\clearpage
%\newpage
\vspace{13pt}\\{\Large{}\textbf{Remerciements}}\vspace{13pt}\\
\'{E}tude réalisée dans le cadre du projet THERMOFLUIDE-RT financé par la Région Ile de France.

%\vspace{13pt}
%{\Large{}\textbf{Références}}

%il est fortement recommandé d'utiliser l'environnement thebibliography

%\bibliography{BIB.bib}{}
\bibliography{bibliographie}{}
\bibliographystyle{ieeetr}

\end{document}